# Beamforming in Two-Way Fixed Gain Amplify-and-Forward Relay Systems with CCI

Trung Q. Duong*, Himal A. Suraweera†, Hans-Jürgen Zepernick*, and Chau Yuen†

*Blekinge Institute of Technology, Sweden (e-mail: {quang.trung.duong, hans-jurgen.zepernick}@bth.se)
†Singapore University of Technology and Design, Singapore (e-mail: {himalsuraweera, yuenchau}@sutd.edu.sg)

*Abstract*—We analyze the outage performance of a two-way fixed gain amplify-and-forward (AF) relay system with beamforming, arbitrary antenna correlation, and co-channel interference (CCI). Assuming CCI at the relay, we derive the exact individual user outage probability in closed-form. Additionally, while neglecting CCI, we also investigate the system outage probability of the considered network, which is declared if any of the two users is in transmission outage. Our results indicate that in this system, the position of the relay plays an important role in determining the user as well as the system outage probability via such parameters as signal-to-noise imbalance, antenna configuration, spatial correlation, and CCI power. To render further insights into the effect of antenna correlation and CCI on the diversity and array gains, an asymptotic expression which tightly converges to exact results is also derived.

## I. INTRODUCTION

The commonly assumed one-way relaying protocol in the literature is limited in system throughput since two time slots are required per single transmission. The loss of system throughput can be recovered by exploiting the concept of two-way relay transmission [1]. Consequently, two-way relaying has attracted a lot of interest, e.g., [2]–[4].

In addition, multiple antenna deployment in fixed gain amplify-and-forward (AF) relay systems can bring further gains at a low practical implementation complexity. In particular, hop-by-hop beamforming is a good technique to realize the benefits of deploying multiple antennas [5]. However, due to space limitation in transmitters and receivers, antenna correlation can degrade the system performance. For one-way beamforming transmission, by considering antenna correlation at the transceiver, the performance of a dual-hop fixed gain AF relay network over Rayleigh fading channels has been investigated in [5], [6].

Besides antenna correlation, co-channel interference (CCI) presenting in cellular systems is another significant impairment that can degrade the system performance. Therefore, a substantial number of research works [7]–[10] have been carried out to investigate the effect of CCI on the performance of fixed gain AF relaying. The outage probability of a fixed gain relay system with CCI at the destination has been reported in [7]. In [8], the effect of multiple Rayleigh interferers at the relay has been quantified. In [9] and [10], the combined effect of CCI at both the relay and destination on the outage probability has been investigated. However, the systems considered in [9], [10] are limited to single antenna model.

In this paper, we consider a two-way relay system in which the source terminals are deployed with multiple antennas. Communication between the two users is facilitated using a single fixed gain AF relay. Furthermore, due to the use of multiple antennas, we consider beamforming and maximal ratio combining for coherent detection. To the best of the authors' knowledge, the effect of CCI on the performance of single or multiple antenna two-way AF relay systems has not been addressed.

In particular, we derive a closed-form expression for the outage probability by considering antenna correlation at user terminals and CCI at the relay. In order to gain further insights, we also develop high SNR outage probability expressions. The asymptotic expression reveals that in a fixed and relatively low interference environment, antenna correlation of a full rank correlation matrix has no impact on the diversity gain, which is equal to the minimum number of antennas equipped at the two sources. In order to comprehensively evaluate the performance of two-way communication further, we analyze the system outage probability, where an outage event is declared when any of the two sources is in outage. Thus it is a measure of the overall Quality-of-Service (QoS) that the system can offer for two-way communication. We note that the system outage has not been widely understood even for simple single antenna two-way relay systems with no CCI effect. The obtained result highlights the significant role of relay placement on the outage performance.

*Notation:* The superscripts $T$ and $\dagger$ stand for the transpose and transpose conjugate. $\|\boldsymbol{y}\|_\mathrm{F}$ denotes Frobenius norm of the vector $\boldsymbol{y}$. $\mathbb{E}_X\{.\}$ is the expectation operator w.r.t. the random variable (RV), $X$. $\Gamma(n)$ is the gamma function [11, Eq. (8.310.1)], $\Gamma(a,x)$ is the incomplete gamma function [11, Eq. (8.350.2)], $\psi(x)$ is Euler psi function [11, Eq. (8.360.1)], $\mathrm{Ei}_n(x) = \int_1^\infty e^{-xt}/t^n dt$ is the generalized exponential integral function, and $\mathcal{K}_n(.)$ is the $n$th-order modified Bessel function of the second kind [11, Eq. (8.432.6)].

## II. SYSTEM AND CHANNEL MODEL

Consider a dual-hop two-way AF relay network where two sources $\mathsf{S}_1$ and $\mathsf{S}_2$ equipped with $N_1$ and $N_2$ antennas, respectively, communicate using a single antenna AF relay, $\mathsf{R}$. The antenna arrays at $\mathsf{S}_1$ and $\mathsf{S}_2$ experience spatial antenna correlation, which can be characterized by two spatial correlation matrices $\boldsymbol{\Xi}_1$ and $\boldsymbol{\Xi}_2$, respectively. The communication

This research is partly supported by the Singapore University Technology and Design (Grant No. SUTD-ZJU/RES/02/2011).

between $S_1$ and $S_2$ occurs in two hops. In the first hop, $S_1$ and $S_2$ simultaneously transmit two messages $s_1$ and $s_2$ to R. Assume that $S_1$ and $S_2$ have the perfect knowledge for channel state information (CSI) of the links to R, the transmit beamforming can be performed by multiplying $s_1$ and $s_2$ with weighting vectors $\mathbf{w}_{T1}$ and $\mathbf{w}_{T2}$, respectively. The signal received at R, corrupted by $L$ interferes, is given by

$$y_R = \boldsymbol{h}_1^T \boldsymbol{\Xi}_1^{1/2} \mathbf{w}_{T1} s_1 + \boldsymbol{h}_2 \boldsymbol{\Xi}_2^{1/2} \mathbf{w}_{T2} s_2 + \sum_{\ell=1}^{L} g_\ell x_\ell + z_R, \quad (1)$$

where $\boldsymbol{h}_n = [h_{n,1}, h_{n,1}, \ldots, h_{n,N_n}]^T$, for $n \in \{1, 2\}$, is the $N_n \times 1$ channel vector from $S_n$ to R with average channel power $\Omega_n$, $z_R$ is additive white Gaussian noise (AWGN) with zero mean and variance $N_0$. In (1), $x_\ell$ and $g_\ell$ are the interfering signal and channel coefficient from the $\ell$-th interferer to R, where $\mathbb{E}\{|x_\ell|^2\} = \mathcal{P}_\ell$ and $\mathbb{E}\{|g_\ell|^2\} = \Omega_{3\ell}$ for $\ell = 1, 2, \ldots, L$.

In the second time-slot, R amplifies $y_R$ with the fixed gain $\mathcal{G}$ before forwarding the resulting signal to both $S_1$ and $S_2$. At each source, e.g., $S_2$, the $N_2 \times 1$ received vector signal $\boldsymbol{y}_{S_2}$ is given by

$$\begin{aligned}\boldsymbol{y}_{S_2} &= \mathcal{G} \boldsymbol{\Xi}_2^{1/2} \boldsymbol{h}_2 \boldsymbol{\Xi}_1^{1/2} \boldsymbol{h}_1 \mathbf{w}_{T1} s_1 + \mathcal{G} \boldsymbol{\Xi}_2^{1/2} \boldsymbol{h}_2 \boldsymbol{h}_2 \boldsymbol{\Xi}_2^{1/2} \mathbf{w}_{T2} s_2 \\ &+ \mathcal{G} \boldsymbol{\Xi}_2^{1/2} \boldsymbol{h}_2 \sum_{\ell=1}^{L} g_\ell x_\ell + \mathcal{G} \boldsymbol{\Xi}_2^{1/2} \boldsymbol{h}_2 z_R + \boldsymbol{z}_2,\end{aligned} \quad (2)$$

where $\boldsymbol{z}_2$ is the $N_2 \times 1$ AWGN vector with zero mean and variance $N_0$. Since each source node knows its own transmitted signal, $S_2$ can remove the self-interference part, i.e., the second term in (2), which requires the knowledge of $\bar{\gamma}_{I_\ell} = \mathcal{P}_\ell / N_0 \mathbb{E}\{|g_\ell|^2\}^1$, $\ell = 1, \ldots, L$. We then multiply the received signal with an $1 \times N_2$ received weighting vector $\mathbf{w}_{R2}$ to yield

$$\begin{aligned}\tilde{y}_{S_2} &= \mathbf{w}_{R2} \mathcal{G} \boldsymbol{\Xi}_2^{1/2} \boldsymbol{h}_2 \boldsymbol{\Xi}_1^{1/2} \boldsymbol{h}_1 \mathbf{w}_{T1} s_1 + \\ &+ \mathbf{w}_{R2} \mathcal{G} \boldsymbol{\Xi}_2^{1/2} \boldsymbol{h}_2 \sum_{\ell=1}^{L} g_\ell x_\ell + \mathbf{w}_{R2} \mathcal{G} \boldsymbol{\Xi}_2^{1/2} \boldsymbol{h}_2 z_R + \mathbf{w}_{R2} \boldsymbol{z}_2.\end{aligned} \quad (3)$$

According to the principle of hop-by-hop beamforming [8], to maximize the signal-to-interference plus noise ratio (SINR), transmit weighting vector at $S_1$, i.e., $\mathbf{w}_{T1}$, and receive weighting vector at $S_2$, i.e., $\mathbf{w}_{R2}$, are chosen as follows:

$$\mathbf{w}_{T1} = (\boldsymbol{\Xi}_1^{\frac{1}{2}} \boldsymbol{h}_1) / \left\| \boldsymbol{\Xi}_1^{\frac{1}{2}} \boldsymbol{h}_1 \right\|_F, \quad \mathbf{w}_{R2} = (\boldsymbol{\Xi}_2^{\frac{1}{2}} \boldsymbol{h}_2) / \left\| \boldsymbol{\Xi}_2^{\frac{1}{2}} \boldsymbol{h}_2 \right\|_F. \quad (4)$$

From (3) and (4), the SINR at $S_2$ is expressed as

$$\gamma_{S_2} = \frac{\mathbb{E}\{|s_1|^2\} \left\| \boldsymbol{\Xi}_1^{1/2} \boldsymbol{h}_1 \right\|_F^2 \left\| \boldsymbol{\Xi}_2^{1/2} \boldsymbol{h}_2 \right\|_F^2}{\left\| \boldsymbol{\Xi}_2^{1/2} \boldsymbol{h}_2 \right\|_F^2 \left( \sum_{\ell=1}^{L} \mathcal{P}_\ell |g_\ell|^2 + N_0 \right) + N_0 / \mathcal{G}^2}. \quad (5)$$

Without loss of generality, we assume that the two sources transmit the same amount of power, i.e., $\mathbb{E}\{|s_1|^2\} =$

---

[1] R can convey these average values to $S_1$ and $S_2$ and update them periodically using a low rate feedback channel.

$\mathbb{E}\{|s_2|^2\} = \mathcal{P}_s$. To maintain the average transmit power at the relay, the amplifying gain $\mathcal{G}$ is determined as

$$\frac{1}{\mathcal{G}^2} = \mathbb{E}\left\{ \left\| \boldsymbol{\Xi}_1^{\frac{1}{2}} \boldsymbol{h}_1 \right\|_F^2 + \left\| \boldsymbol{\Xi}_2^{\frac{1}{2}} \boldsymbol{h}_2 \right\|_F^2 + \sum_{\ell=1}^{L} \frac{\mathcal{P}_\ell}{\mathcal{P}_s} |g_\ell|^2 + \frac{N_0}{\mathcal{P}_s} \right\}. \quad (6)$$

To simplify the notation, we denote $\gamma_n = \bar{\gamma} \left\| \boldsymbol{\Xi}_n^{1/2} \boldsymbol{h}_n \right\|_F^2$, for $n \in \{1, 2\}$, and $\gamma_3 = \sum_{\ell=1}^{L} \mathcal{P}_\ell / N_0 |g_\ell|^2$, with $\bar{\gamma} = \mathcal{P}_s / N_0$. The instantaneous SINR at $S_n$, for $n \in \{1, 2\}$, can be rewritten as

$$\gamma_{S_n} = \frac{\gamma_1 \gamma_2}{\gamma_n (\gamma_3 + 1) + \mathcal{C}}, \quad (7)$$

where $\mathcal{C}$ is a constant given by $\mathcal{C} = \mathcal{P}_s / (N_0 \mathcal{G}^2)$, which will be derived in the sequel.

It is important to obtain the statistical characteristics of RV $\gamma_n$, for $n \in \{1, 2\}$, given in (7). Without loss of generality, assuming that the correlation matrix $\boldsymbol{\Xi}_n$ has $Q_n$ distinct non-zero eigenvalues $\lambda_{n1}, \lambda_{n2}, \ldots, \lambda_{nQ_n}$ with multiplicities $\alpha_{n1}, \alpha_{n2}, \ldots, \alpha_{nQ_n}$, respectively, the probability density function (PDF) and cumulative distribution function (CDF) of $\gamma_n$ can be formulated as

$$f_{\gamma_n}(\gamma) = \sum_{i=1}^{Q_n} \sum_{j=1}^{\alpha_{ni}} \frac{\vartheta_{nij} \gamma^{j-1}}{\Gamma(j)(\bar{\gamma} \chi_{ni})^j} e^{-\frac{\gamma}{\bar{\gamma} \chi_{ni}}}, \quad (8)$$

$$F_{\gamma_n}(\gamma) = 1 - \sum_{i=1}^{Q_n} \sum_{j=1}^{\alpha_{ni}} \sum_{k=0}^{j-1} \frac{\vartheta_{nij}}{k!} \left( \frac{\gamma}{\bar{\gamma} \chi_{ni}} \right)^k e^{-\frac{\gamma}{\bar{\gamma} \chi_{ni}}}, \quad (9)$$

where $\chi_{ni} = \lambda_{ni} \Omega_n$ and the expansion coefficient $\vartheta_{nij}$ is defined as

$$\vartheta_{nij} = \frac{(\chi_{nij})^{\alpha_{ni}-j}}{(\alpha_{ni}-j)!} \frac{d^{\alpha_{ni}-j}}{dt^{\alpha_{ni}-j}} \left[ \prod_{l=1, l \neq i}^{Q_n} (t + \chi_{nij})^{-\alpha_{ni}} \right] \Bigg|_{t=-\chi_{nij}}.$$

When the channels are independent, i.e., $\boldsymbol{\Xi}_n$ is identity matrix, we have $\vartheta_{nij} = 1$ for $i = 1$, $j = 1, 2, \ldots, N_n$ and $\vartheta_{nij} = 0$ for $i = 1$, $j = 1, 2, \ldots, N_n - 1$. When the correlation matrix follows an exponential model, i.e., all the eigenvalues $\lambda_{n1}, \lambda_{n2}, \ldots, \lambda_{nN_n}$ are distinct, we have $\vartheta_{ni} = \chi_{ni}^{N_n-1} / \prod_{l=1, l \neq i}^{N_n} (\chi_{ni} - \chi_{nl})$.

Since $\gamma_3$ is the sum of $L$ exponentially distributed RVs, its PDF is easily given by $f_{\gamma_3}(\gamma) = \sum_{\ell=1}^{L} \beta_\ell \exp\left(\frac{\gamma}{\bar{\gamma}_{I_\ell}}\right)$, where $\beta_\ell = \bar{\gamma}_{I_\ell}^{-1} \prod_{k=1, k \neq \ell} (1 - \bar{\gamma}_{I_k} / \bar{\gamma}_{I_\ell})^{-1}$. Using the above correlation model, constant $\mathcal{C}$ is obtained as

$$\mathcal{C} = \bar{\gamma} \sum_{i=1}^{Q_1} \sum_{j=1}^{\alpha_{1i}} \vartheta_{1ij} j \chi_{1i} + \bar{\gamma} \sum_{r=1}^{Q_2} \sum_{t=1}^{\alpha_{2r}} \vartheta_{2rt} t \chi_{2r} + \sum_{\ell=1}^{L} \frac{\beta_\ell}{\bar{\gamma}_{I_\ell}^2} + 1. \quad (10)$$

## III. OUTAGE PROBABILITY ANALYSIS

The outage event occurs when the instantaneous SINR falls below a predefined threshold. In this work, we consider the outage probability in two cases: i) the user outage probability at $S_1$ or $S_2$ and ii) system outage probability declared when the minimum SINR between $S_1$ and $S_2$ is below a threshold.

## A. User Outage Probability

In this particular case, the outage probability is defined as the probability that $S_1$ or $S_2$ is in outage, i.e., $P_{\text{out}} = \Pr(\gamma_{S_n} < \gamma_{\text{th}}) = F_{\gamma_{S_n}}(P_{\text{out}})$.

*1) Exact Outage Probability:* Consider the outage probability for $S_2$. The exact CDF of $\gamma_{S_2}$ is given by (see Appendix A for proof)

$$P_{\text{out}} = 1 - 2 \sum_{i=1}^{Q_1} \sum_{j=1}^{\alpha_{1i}} \sum_{k=0}^{j-1} \frac{\vartheta_{1ij}}{k!(\chi_{1i}\bar{\gamma})^k} \sum_{l=0}^{k} \binom{k}{l} \gamma_{\text{th}}^k \mathcal{C}^{k-l}$$
$$\times e^{-\gamma_{\text{th}}/(\bar{\gamma}\chi_{1i})} \sum_{r=1}^{Q_2} \sum_{t=1}^{\alpha_{2r}} \frac{\vartheta_{2rt}}{\Gamma(t)(\chi_{2r}\bar{\gamma})^t} \mathcal{K}_{l+t-k}\left(2\sqrt{\frac{\gamma_{\text{th}}\mathcal{C}}{\bar{\gamma}^2 \chi_{1i}\chi_{2r}}}\right)$$
$$\times \left(\frac{\gamma_{\text{th}}\mathcal{C}\chi_{2r}}{\chi_{1i}}\right)^{\frac{l+t-k}{2}} \sum_{s=0}^{l} \frac{l!}{(l-s)!} \sum_{\ell=1}^{L} \beta_\ell \left(\frac{\gamma_{\text{th}}}{\bar{\gamma}\chi_{1i}} + \frac{1}{\bar{\gamma}_{I_\ell}}\right)^{-s-1}. \quad (11)$$

When a particular channel correlation model is adopted, we can further simplify (12). For e.g., in the case of exponential correlation, (11) reduces to

$$P_{\text{out}} = 1 - 2\sum_{i=1}^{N_1} \vartheta_{1i} e^{-\frac{\gamma_{\text{th}}}{\bar{\gamma}\chi_{1i}}} \sum_{r=1}^{N_2} \frac{\vartheta_{2r}}{\chi_{2r}\bar{\gamma}} \sqrt{\frac{\gamma_{\text{th}}\mathcal{C}\chi_{2r}}{\chi_{1i}}}$$
$$\times \mathcal{K}_1\left(2\sqrt{\frac{\gamma_{\text{th}}\mathcal{C}}{\bar{\gamma}^2\chi_{1i}\chi_{2r}}}\right) \sum_{\ell=1}^{L} \beta_\ell \left(\frac{\gamma_{\text{th}}}{\bar{\gamma}\chi_{1i}} + \frac{1}{\bar{\gamma}_{I_\ell}}\right)^{-1}. \quad (12)$$

For the case of independent fading, (11) simplifies to

$$P_{\text{out}} = 1 - 2\sum_{i=0}^{N_1-1} \frac{e^{-\frac{\gamma_{\text{th}}}{\bar{\gamma}\Omega_1}}}{i!(\bar{\gamma}\Omega_1)^i} \sum_{l=0}^{i} \frac{\binom{i}{l} \mathcal{C}^{i-l}\gamma_{\text{th}}^i}{\Gamma(N_2)(\bar{\gamma}\Omega_2)^{N_2}} \left(\frac{\gamma_{\text{th}}\mathcal{C}\Omega_2}{\Omega_1}\right)^{\frac{N_2+l-i}{2}}$$
$$\times \mathcal{K}_{N_2+l-i}\left(2\sqrt{\frac{\gamma_{\text{th}}\mathcal{C}}{\bar{\gamma}^2\Omega_1\Omega_2}}\right) \sum_{s=0}^{l} \frac{l!}{(l-s)!} \sum_{\ell=1}^{L} \beta_\ell \left(\frac{\gamma_{\text{th}}}{\bar{\gamma}\Omega_1} + \frac{1}{\bar{\gamma}_{I_\ell}}\right)^{-1}. \quad (13)$$

*2) Outage Probability at High SNR:* To provide additional insights into the behavior of the outage probability and to investigate the diversity order and the array gain of the system, we now present an asymptotic result. In the high signal-to-noise ratio (SNR) regime, i.e., $\bar{\gamma} \to \infty$, we can express $\mathcal{C} \approx \varrho\bar{\gamma}$ where $\varrho$ is a constant, which results in (see Appendix B for proof)

$$P_{\text{out}}^\infty = \begin{cases} c_{(N_1)} \left(\frac{\gamma_{\text{th}}}{\bar{\gamma}}\right)^{N_1} & \text{if } N_1 < N_2 \\ c_{(N_3)} \left(\frac{\gamma_{\text{th}}}{\bar{\gamma}}\right)^{N_3} & \text{if } N_1 = N_2 = N_3 \\ c_{(N_2)} \left(\frac{\gamma_{\text{th}}}{\bar{\gamma}}\right)^{N_2} & \text{if } N_1 > N_2 \end{cases}, \quad (14)$$

where constant $c_{(\theta)}$, for $\theta = \min(N_1, N_2)$, is written as

$$c_{(\theta)} = \sum_{i=1}^{Q_1} \sum_{j=1}^{\alpha_{1i}} \sum_{k=0}^{j-1} \frac{\vartheta_{1ij}}{k!} \sum_{l=0}^{k} \binom{k}{l} \sum_{r=1}^{Q_2} \sum_{t=1}^{\alpha_{2r}} \frac{\vartheta_{2rt}}{\Gamma(t)} \sum_{s=0}^{l} \frac{l!}{(l-s)!}$$
$$\times \sum_{\ell=1}^{L} \beta_\ell \left(\frac{\gamma}{\bar{\gamma}\chi_{1i}} + \frac{1}{\bar{\gamma}_{I_\ell}}\right)^{-s-1} \left(\frac{1}{\chi_{1i}}\right)^\theta \Phi(l,t,k). \quad (15)$$

Recall that $\bar{\gamma} = \mathcal{P}_s/N_0$ and $\bar{\gamma}_{I_\ell} = \mathcal{P}_\ell/N_0$. In (14), we have assumed that the interference-to-noise ratio (INR), $\bar{\gamma}_{I_\ell}$, is low and fixed ($\bar{\gamma}_{I_\ell}$ does not vary when the SNR is increased). On the other hand, when a symmetric network is assumed, such that the interfering terminals transmit with the same power characteristics as the useful terminals, (implying that the INR tends to infinity when the SNR tends to infinity) e.g., [8], [12], the diversity order becomes zero regardless of the use of multiple antennas at $S_1$ and $S_2$. Result $\Phi(l,t,k)$ in (15) is defined according to the relationship of the running indices $l$, $t$, and $k$ as follows:

- If $l + t - k > 0$, we have

$$\Phi(l,t,k) = \sum_{w=0}^{\min(l+t-k-1,\theta-k)} \frac{(-1)^{\theta-k+1}(l+t-k-w-1)!}{w!(\theta-k-w)!}$$
$$\times \left(\frac{\varrho}{\chi_{2r}}\right)^{k-l+w} + \sum_{w=0}^{\theta-l-t} \frac{(-1)^{\theta-k-w}/(\theta-t-l-w)!}{w!(l+t-k+w)!}$$
$$\times \left(\frac{\varrho}{\chi_{2r}}\right)^{t+w} \left[\log\left(\frac{\varrho\gamma_{\text{th}}}{\bar{\gamma}\chi_{1i}\chi_{2r}}\right) - \psi(w+1)\right.$$
$$\left. - \psi(l+t-k+w+1)\right]. \quad (16)$$

- If $l + t - k = 0$, we have $\Phi(l,t,k) = \Phi(t,k)$, shown as

$$\Phi(t,k) = \sum_{w=0}^{\theta-k} \frac{\log\left(\frac{\varrho\gamma_{\text{th}}}{\bar{\gamma}\chi_{1i}\chi_{2r}}\right) - 2\psi(w+1)}{(-1)^{k+w-\theta}w!w!(\theta-k-w)!\chi_{2r}^{t+w}}. \quad (17)$$

- If $l + t - k < 0$, we have

$$\Phi(l,t,k) = \sum_{w=0}^{\min(k-l-t-1,\theta-t-l)} \frac{(-1)^{\theta-l-t+1}(k-l-t+w-1)!}{w!(\theta-t-l-w)!}$$
$$\times \left(\frac{\varrho}{\chi_{2r}}\right)^{w+t} + \sum_{w=0}^{\theta-k} \frac{(-1)^{\theta-t-l-w}}{w!(k-l-t+w)!(\theta-k-w)!}$$
$$\times \left(\frac{\varrho}{\chi_{2r}}\right)^{k+w-l} \left[\log\left(\frac{\varrho\gamma_{\text{th}}}{\bar{\gamma}\chi_{1i}\chi_{2r}}\right) - \psi(w+1)\right.$$
$$\left. - \psi(k-l-t+w+1)\right]. \quad (18)$$

For an exponential correlation model, $c_{(\theta)}$ given in (15) simplifies to

$$c_{(\theta)} = \sum_{i=1}^{N_1} \sum_{r=1}^{N_2} \vartheta_{1i}\vartheta_{2r} \sum_{\ell=1}^{L} \frac{\beta_\ell}{\chi_{1i}^\theta} \left(\frac{\gamma}{\bar{\gamma}\chi_{1i}} + \frac{1}{\bar{\gamma}_{I_\ell}}\right)^{-1} \Phi_1, \quad (19)$$

where

$$\Phi_1 = \sum_{w=0}^{\theta-1} \frac{(-1)^\theta \left[\ln\left(\frac{\gamma_{\text{th}}\mathcal{C}}{\bar{\gamma}\chi_{1i}\chi_{2r}}\right) - \psi(w+1) - \psi(w+2)\right]}{w!(w+1)!(\theta-w-1)!}$$
$$\times \left(\frac{\varrho}{\chi_{2r}}\right)^{w+1} + \frac{(-1)^{\theta+1}}{\theta!}. \quad (20)$$

From (14), it can be observed that when the channel correlation matrices are of full-rank, the diversity order is equal to the minimum of antennas at $S_1$ and $S_2$, i.e., $\min(N_1, N_2)$, which is the maximum achievable diversity gain, and that correlation does not affect the diversity order.

## B. System Outage Probability Analysis

The two-way relaying concept considers information exchange between $S_1$ and $S_2$. Therefore, in some applications, successful transmission is declared only when both $S_1$ and $S_2$ in operation. In other words, the considered system is suspended if any of $S_1$ and $S_2$ is in outage [4]. We define the system outage probability as

$$P_{\text{out}} = \Pr\left(\min(\gamma_{S_1}, \gamma_{S_2}) < \gamma_{\text{th}}\right). \quad (21)$$

Since the two RVs $\gamma_{S_1}$ and $\gamma_{S_2}$ are dependent, in this case the mathematical derivation is much involved. Therefore, for the mathematical tractability, we omit the effect of CCI. The instantaneous SINRs given in (7) are now rewritten as $\gamma_{S_1} = \frac{\gamma_1 \gamma_2}{\gamma_1 + \mathcal{C}}$ for $S_1$ and $\gamma_{S_2} = \frac{\gamma_1 \gamma_2}{\gamma_2 + \mathcal{C}}$ for $S_2$, which leads to

$$P_{\text{out}} = \Pr\left[\min(\gamma_{S_1}, \gamma_{S_2}) < \gamma_{\text{th}}\right]$$
$$= \underbrace{\Pr\left(\gamma_{S_1} < \gamma_{\text{th}}, \gamma_{S_1} < \gamma_{S_2}\right)}_{\mathcal{I}_1} + \underbrace{\Pr\left(\gamma_{S_2} < \gamma_{\text{th}}, \gamma_{S_2} < \gamma_{S_1}\right)}_{\mathcal{I}_2}. \quad (22)$$

Because of the symmetry between $\mathcal{I}_1$ and $\mathcal{I}_2$ in (22), we concentrate on $\mathcal{I}_1$. Since the condition $\gamma_{S_1} < \gamma_{S_2}$ is equivalent to $\gamma_1 > \gamma_2$, we obtain

$$\mathcal{I}_1 = \underbrace{\int_\epsilon^\infty f_{\gamma_1}(\gamma_1) \int_0^{\gamma_{\text{th}} + \frac{\gamma_{\text{th}} \mathcal{C}}{\gamma_1}} f_{\gamma_2}(\gamma_2) \, d\gamma_2 d\gamma_1}_{\mathcal{J}_1}$$
$$+ \underbrace{\int_0^\epsilon f_{\gamma_1}(\gamma_1) \int_0^{\gamma_1} f_{\gamma_2}(\gamma_2) \, d\gamma_2 d\gamma_1}_{\mathcal{J}_2}, \quad (23)$$

where $\epsilon = \frac{1}{2}\left(\gamma_{\text{th}} + \sqrt{\gamma_{\text{th}}^2 + 4\gamma_{\text{th}}\mathcal{C}}\right)$ is the positive root of the quadratic equation $\gamma_1^2 - \gamma_{\text{th}}\gamma_1 - \gamma_{\text{th}}\mathcal{C} = 0$. The exact expression for $\mathcal{I}_1$ is given by (see Appendix C for detailed proof)

$$\mathcal{I}_1 = 1 - \sum_{i=1}^{Q_1} \sum_{j=1}^{\alpha_{1i}} \frac{\vartheta_{1ij}}{\Gamma(j)(\chi_{1i}\bar{\gamma})^j} \sum_{r=1}^{Q_2} \sum_{t=1}^{\alpha_{2r}} \sum_{k=0}^{t-1} \frac{\vartheta_{2rt}}{k!} \left(\frac{\gamma_{\text{th}}}{\bar{\gamma}\chi_{2r}}\right)^k$$
$$\times \left\{ e^{-\gamma_{\text{th}}/(\bar{\gamma}\chi_{2r})} \sum_{l=0}^k \binom{k}{l} \mathcal{C}^{k-l} \sum_{s=0}^\infty \left(\frac{-\gamma_{\text{th}}\mathcal{C}}{\bar{\gamma}\chi_{2r}}\right)^s \frac{\epsilon^{j+l-s-k}}{s!} \right.$$
$$\times \text{Ei}_{s+k-j-l+1}\left(\frac{\epsilon}{\bar{\gamma}\chi_{1i}}\right) + \left(\frac{1}{\bar{\gamma}\chi_{1i}} + \frac{1}{\bar{\gamma}\chi_{2r}}\right)^{-j-k}$$
$$\left. \times \left[\Gamma(j+k) - \Gamma\left(j+k, \frac{\epsilon}{\bar{\gamma}\chi_{1i}} + \frac{\epsilon}{\bar{\gamma}\chi_{2r}}\right)\right] \right\}. \quad (24)$$

Similarly, $\mathcal{I}_2$ is given by

$$\mathcal{I}_2 = 1 - \sum_{i=1}^{Q_2} \sum_{j=1}^{\alpha_{2i}} \frac{\vartheta_{2ij}}{\Gamma(j)(\chi_{2i}\bar{\gamma})^j} \sum_{r=1}^{Q_1} \sum_{t=1}^{\alpha_{1r}} \sum_{k=0}^{t-1} \frac{\vartheta_{1rt}}{k!} \left(\frac{\gamma_{\text{th}}}{\bar{\gamma}\chi_{1r}}\right)^k$$
$$\times \left\{ e^{-\gamma_{\text{th}}/(\bar{\gamma}\chi_{1r})} \sum_{l=0}^k \binom{k}{l} \mathcal{C}^{k-l} \sum_{s=0}^\infty \left(\frac{-\gamma_{\text{th}}\mathcal{C}}{\bar{\gamma}\chi_{1r}}\right)^s \frac{\epsilon^{j+l-s-k}}{s!} \right.$$
$$\times \text{Ei}_{s+k-j-l+1}\left(\frac{\epsilon}{\bar{\gamma}\chi_{2i}}\right) + \left(\frac{1}{\bar{\gamma}\chi_{2i}} + \frac{1}{\bar{\gamma}\chi_{1r}}\right)^{-j-k}$$
$$\left. \times \left[\Gamma(j+k) - \Gamma\left(j+k, \frac{\epsilon}{\bar{\gamma}\chi_{2i}} + \frac{\epsilon}{\bar{\gamma}\chi_{1r}}\right)\right] \right\}. \quad (25)$$

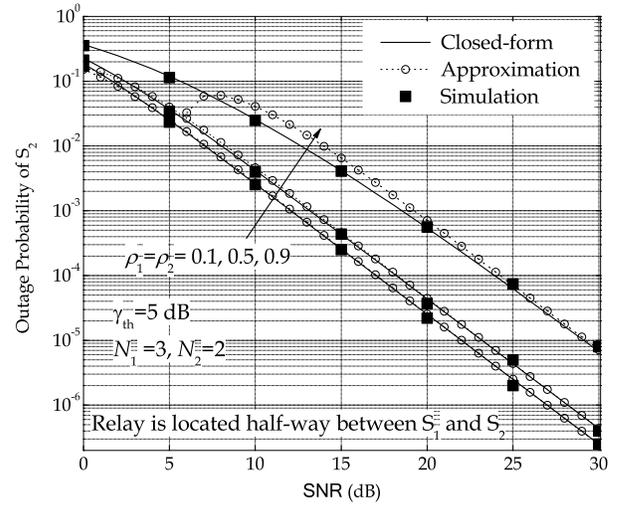

Fig. 1. Outage probability of $S_2$ with different correlation coefficients. Results are shown for $L = 1$ and $\bar{\gamma}_{I_1} = 1$ dB.

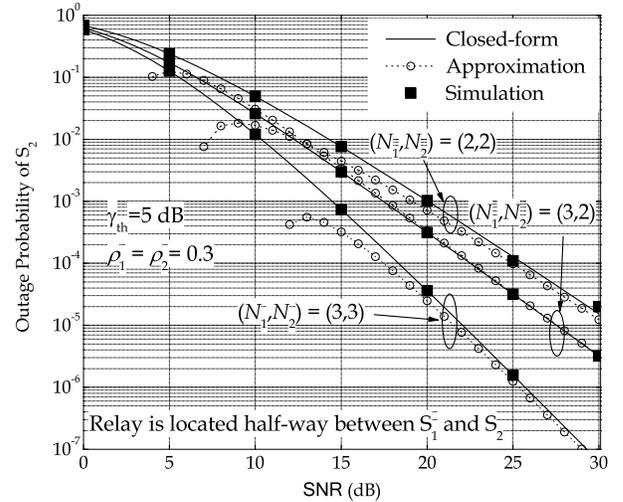

Fig. 2. Outage probability of $S_2$ with different antenna configurations. Results are shown for $L = 3$ and $\{\bar{\gamma}_{I_\ell}\}_{\ell=1}^3 = \{1, 2, 3\}$ dB.

The system outage probability is expressed in the form of one infinite sum which is shown to converge very fast. For example, only five number of terms is required in the summation over index $s$ to achieve the accuracy to the degree of eight decimals.

## IV. NUMERICAL RESULTS AND DISCUSSION

In this section, we provide numerical results to validate the above analysis. Here we apply an exponential-decay model for the path loss. Specifically, assume that the distance between $S_1$ and $S_2$ is equal to $d$, we have the corresponding channel mean power $\Omega_0 \sim d^{-\mu}$. Then, $\Omega_1 = \kappa^{-\mu}\Omega_0$ and $\Omega_2 = (1-\kappa)^{-\mu}\Omega_0$, where $\kappa$ stands for the fraction of the distance from $S_1$ to R over the distance from $S_1$ to $S_2$. For example, when the relay is located in the middle between $S_1$ and $S_2$, we have $\kappa = 0.5$. Moreover, an exponential correlation model is used where the correlation coefficient between the $i$-th and $j$-th antennas of $S_n$ is given by $\rho_{n_{i,j}} = \rho_n^{|i-j|}$ with $i, j = 1, 2, \ldots, N_n$ and $n = 1, 2$. In all examples, we set $\gamma_{th} = 5$ dB and $\Omega_0 = 1$. Unless otherwise stated, R is located half-way between $S_1$ and $S_2$, leading to $\Omega_1 = \Omega_2 = 16$.

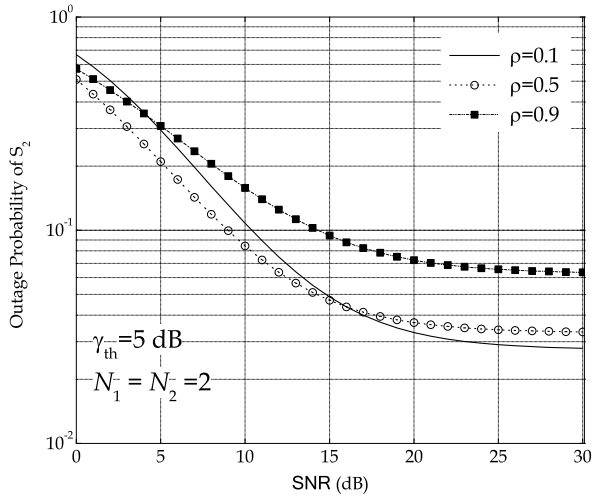

Fig. 3. Effect of CCI on the outage probability of $S_2$ when $\bar{\gamma}_{I_\ell} = \bar{\gamma}$.

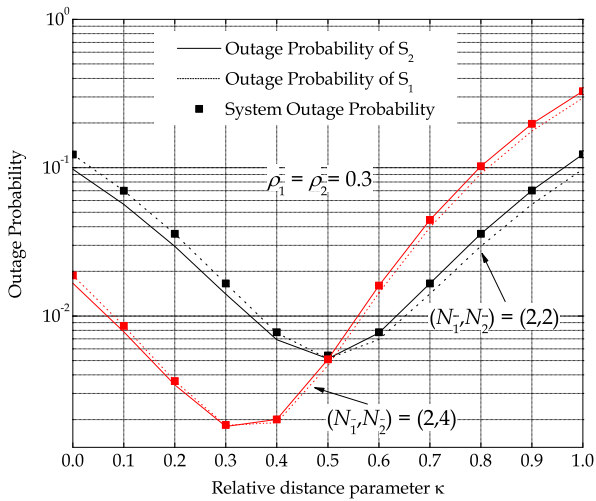

Fig. 4. User outage probability of $S_1$, $S_2$, and system outage probability versus the relative position of relay.

Fig. 1 shows the outage probability of $S_2$ for the case of $N_1 = 3$, $N_2 = 2$ and a single interferer with $\bar{\gamma}_{I_1} = 1$ dB. As expected, we see that high correlation adversely degrades the outage probability. However, only the array gain is affected by correlation while the diversity gain remains the same since the three curves are parallel as plotted in log-log scale. Moreover, the exact curves plotted from (11) are in excellent agreement with Monte Carlo simulations and the asymptotic curves plotted from (14) converge to the exact curves.

Fig. 2 shows the outage probability for a fixed correlation coefficient, i.e., $\rho_1 = \rho_2 = 0.3$, and $L = 3$ with $\{\bar{\gamma}_{I_\ell}\}_{\ell=1}^3 = \{1, 2, 3\}$ dB. To clearly highlight the effect of diversity, the antenna configuration at $S_1$ and $S_2$ is selected as $(N_1, N_2) = (2, 2), (3, 2), (3, 3)$. We notice that for a fixed $N_1$, increasing $N_2$ yields no additional diversity gain as the two systems $(2, 2)$ and $(3, 2)$ have the same diversity order.

Fig. 3 illustrates the effect of CCI on the system performance by setting $\bar{\gamma}_{I_\ell}$ proportional to the average SNR as $\bar{\gamma}_{I_\ell} = \nu \bar{\gamma}$, where $\nu$ is a fixed scalar. We see that, correlation significantly degrades the outage probability as all the curves exhibit a floor in the high SNR regime. It is interesting to observe that correlation improves the performance in the low SNR region. For low SNR, correlation allows potentially focused power, which is beneficial to the system performance [13]. This explains the observed effect in Fig. 3.

In Fig. 4, in order to show the effect of the relay position, we have plotted the user and system outage probability versus the relative distance parameter $\kappa$ for the symmetric case, e.g., $(N_1, N_2) = (2, 2)$, and asymmetric case, e.g., $(N_1, N_2) = (2, 4)$. When R is located nearby $S_1$, we have $\kappa < 0.5$. Notice that $S_2$ outperforms $S_1$ when R is close to $S_1$ and vice versa, which shows that the first hop channel governs the user outage probability. In the considered system, interference at R is amplified before forwarding to a user, and it has a dominant effect when the first-hop link is strong. The system outage probability coincides with the worst case of $S_1$ and $S_2$, as expected. The best performance for the symmetric case is achieved at $\kappa = 0.5$, while it is interesting to see that in the asymptotic case, the best performance is obtained for $\kappa = 0.3$. This observation shows that when the system is balanced, i.e., symmetric topology, R must be placed in the middle between $S_1$ and $S_2$. However, for the unbalanced case, e.g., $(N_1, N_2) = (2, 4)$, R must be nearby $S_1$. The shift of relay location to $S_1$ compensates for the imbalance of the system model. Specifically, in our example, $S_1$ has less number of antennas, hence, in order to achieve a compromise for $S_1$ and $S_2$, a designer must carefully select the system parameters.

## V. CONCLUSIONS

We have investigated the performance of a two-way fixed gain AF relay system with antenna correlation and CCI. In order to analyze the effects of these important practical impairments on the user and system outage probability, we derived the exact closed-form expressions. Moreover, asymptotic result providing further insights into array gain and diversity order was also obtained. We see that multiple antenna deployment is an attractive solution to improve the outage performance when the relay system is affected by low level of CCI. However, if the CCI effect is more severe, the performance significantly deteriorates. As such, multi-antenna interference cancelation schemes could be additionally implemented and considered for future work.

## APPENDIX A

The CDF of $\gamma_{S_2}$ is written as

$$F_{\gamma_{S_2}}(\gamma) = \int_0^\infty \int_0^\infty F_{\gamma_1}\left(\gamma(\gamma_3+1) + \frac{\mathcal{C}\gamma}{\gamma_2}\right) \\ \times f_{\gamma_2}(\gamma_2) f_{\gamma_3}(\gamma_3) d\gamma_2 d\gamma_3. \quad (26)$$

By substituting (8), (9) into (26) and applying the binomial theorem, we obtain

$$F_{\gamma_{S_2}}(\gamma) = 1 - \sum_{i=1}^{Q_1}\sum_{j=1}^{\alpha_{1i}}\sum_{k=0}^{j-1} \frac{\vartheta_{1ij}}{k!(\chi_{1i}\bar{\gamma})^k}\sum_{l=0}^{k}\binom{k}{l}\gamma^k\mathcal{C}^{k-l}$$
$$\times e^{-\gamma/(\bar{\gamma}\chi_{1i})}\sum_{r=1}^{Q_2}\sum_{t=1}^{\alpha_{2r}}\frac{\vartheta_{2rt}}{\Gamma(t)(\chi_{2r}\bar{\gamma})^t}\sum_{\ell=1}^{L}\beta_\ell$$
$$\times \int_0^\infty\int_0^\infty \gamma_2^{l-k+t-1}\exp\left(-\frac{\mathcal{C}\gamma}{\bar{\gamma}\chi_{1i}\gamma_2} - \frac{\gamma_2}{\bar{\gamma}\chi_{2r}}\right)$$
$$\times (\gamma_3+1)^l \exp\left(-\frac{\gamma\gamma_3}{\bar{\gamma}\chi_{1i}} - \frac{\gamma_3}{\bar{\gamma}_{I_\ell}}\right)d\gamma_2 d\gamma_3. \quad (27)$$

To solve the above double integral, we first expand $(\gamma_3+1)^l$ into a finite sum by using binomial theorem [11, Eq. (1.111)], and then use [11, Eq. (4.471.9)]. After some simplifications, we obtain (11).

## APPENDIX B

Due to the space limit, we will briefly introduce the approach to calculate the asymptotic result. First, we expand the exponential term $e^{\frac{-\gamma_{\text{th}}}{\bar{\gamma}\chi_{1i}}}$ into the infinite sum using Taylor expansion. Second, we employ series representation for Bessel function as

$$\mathcal{K}_\nu(z) = \sum_{w=0}^{\nu-1}\frac{\Gamma(\nu-w)}{\Gamma(w+1)}\frac{(-1)^w}{2}\left(\frac{z}{2}\right)^{-\nu+2w} + $$
$$(-1)^{\nu+1}\sum_{w=0}^{\infty}\frac{\left(\frac{z}{2}\right)^{\nu+2w}\left[\ln(\frac{z}{2}) - \frac{\psi(w+1)}{2} - \frac{\psi(\nu+w+1)}{2}\right]}{\Gamma(w+1)\Gamma(\nu+w+1)}. \quad (28)$$

Depending on the relationship of the running indices $l$, $t$, and $k$, $\nu = l+t-k$ can be positive, negative, or zero. When $\nu > 0$ we utilize (28). When $\nu < 0$, we first apply $\mathcal{K}_\nu(\cdot) = \mathcal{K}_{-\nu}(\cdot)$ and then (28). For zero value, i.e., $l+t-k = 0$, the following expansion is valid

$$\mathcal{K}_0(z) = -\ln\frac{z}{2}\sum_{w=0}^{\infty}\frac{\left(\frac{z}{2}\right)^{2w}}{(w!)^2} + \sum_{w=0}^{\infty}\frac{z^{2w}}{2^{2w}(w!)^2}\psi(w+1). \quad (29)$$

Then by substituting the partial coefficients $\vartheta_{nij}$, for the terms $\gamma_{\text{th}}^n$, the sum of these terms becomes zero when $n < \theta$. Therefore, the lowest order of the exponent $n$ is equal to $\theta$.

## APPENDIX C

We first consider $\mathcal{J}_1$ given in (23) as

$$\mathcal{J}_1 = \int_\epsilon^\infty f_{\gamma_1}(\gamma_1) F_{\gamma_2}\left(\gamma_{\text{th}} + \frac{\gamma_{\text{th}}\mathcal{C}}{\gamma_1}\right) d\gamma_1. \quad (30)$$

Now, by substituting (8), (9) into (30), and performing some manipulations yields

$$\mathcal{J}_1 = 1 - F_{\gamma_1}(\epsilon) - \sum_{i=1}^{Q_1}\sum_{j=1}^{\alpha_{1i}}\frac{\vartheta_{1ij}}{\Gamma(j)(\chi_{1i}\bar{\gamma})^j}\sum_{r=1}^{Q_2}\sum_{t=1}^{\alpha_{2r}}\sum_{k=0}^{t-1}\frac{\vartheta_{2rt}}{k!}e^{-\frac{\gamma_{\text{th}}}{\bar{\gamma}\chi_{2r}}}$$
$$\times \left(\frac{\gamma_{\text{th}}}{\bar{\gamma}\chi_{2r}}\right)^k\int_\epsilon^\infty \gamma_1^{j-k-1}(\gamma_1+\mathcal{C})^k e^{-\frac{\gamma_1}{\bar{\gamma}\chi_{1i}}}e^{-\frac{\gamma_{\text{th}}\mathcal{C}}{\bar{\gamma}\chi_{2r}\gamma_1}}d\gamma_1. \quad (31)$$

To the best of our knowledge, the integral in (31) has no closed-form solution. To solve this integral, we first apply binomial theorem [11, Eq. (1.111)] for term $(\gamma_1+\mathcal{C})^k$ and using Taylor series representation [11, Eq. (1.211.1)] for term $e^{-\frac{\gamma_{\text{th}}\mathcal{C}}{\bar{\gamma}\chi_{2r}\gamma_1}}$, which results in

$$\mathcal{J}_1 = 1 - F_{\gamma_1}(\epsilon) - \sum_{i=1}^{Q_1}\sum_{j=1}^{\alpha_{1i}}\frac{\vartheta_{1ij}}{\Gamma(j)(\chi_{1i}\bar{\gamma})^j}\sum_{r=1}^{Q_2}\sum_{t=1}^{\alpha_{2r}}\sum_{k=0}^{t-1}\frac{\vartheta_{2rt}}{k!}$$
$$\times \left(\frac{\gamma_{\text{th}}}{\bar{\gamma}\chi_{2r}}\right)^k e^{-\frac{\gamma_{\text{th}}}{\bar{\gamma}\chi_{2r}}}\sum_{l=0}^{k}\binom{k}{l}\mathcal{C}^{k-l}\sum_{s=0}^{\infty}\left(\frac{-\gamma_{\text{th}}\mathcal{C}}{\bar{\gamma}\chi_{2r}}\right)^s \frac{1}{s!}$$
$$\times \int_\epsilon^\infty \gamma_1^{j+l-k-s-1}e^{-\frac{\gamma_1}{\bar{\gamma}\chi_{1i}}}d\gamma_1. \quad (32)$$

The integral representation in (32) can be obtained in the form of the generalized exponential integral function $\text{Ei}_\nu(\cdot)$.

Next, we evaluate $\mathcal{J}_2$ given in (23) by rewriting it as

$$\mathcal{J}_2 = \int_0^\epsilon f_{\gamma_1}(\gamma_1) F_{\gamma_2}(\gamma_1) d\gamma_1$$
$$= F_{\gamma_2}(\epsilon) - \sum_{i=1}^{Q_1}\sum_{j=1}^{\alpha_{1i}}\frac{\vartheta_{1ij}}{\Gamma(j)(\chi_{1i}\bar{\gamma})^j}\sum_{r=1}^{Q_2}\sum_{t=1}^{\alpha_{2r}}\sum_{k=0}^{t-1}\frac{\vartheta_{2rt}}{k!}\frac{1}{(\bar{\gamma}\chi_{2r})^k}$$
$$\times \int_0^\epsilon \gamma_1^{j+k-1}e^{-\frac{\gamma_1}{\bar{\gamma}\chi_1} - \frac{\gamma_1}{\bar{\gamma}\chi_{2r}}}d\gamma_1. \quad (33)$$

The above integral can be solved in the form of gamma function $\Gamma(n)$ and incomplete gamma function $\Gamma(n,x)$ with the help of [11, Eq. (3.381.8)]. Finally, we sum (32) and (33) and with rearrangement to arrive at (24).